\documentclass[aps,prl,twocolumn,reprint,amsmath,amssymb,superscriptaddress,floatfix]{revtex4-1}

\usepackage{graphicx}
\usepackage{multirow}
\usepackage{color}
\usepackage{bm}
\usepackage{hyperref}

\begin{document}

\title{Theory of shift current in Anderson insulator}

\author{Hiroaki~Ishizuka}
\affiliation{Department of Physics, Massachusetts Institute of Technology, 
Cambridge Massachusetts, 02139, USA}
\affiliation{Department of Applied Physics, The University of Tokyo, Hongo, 
Tokyo, 113-8656, Japan}

\author{Naoto~Nagaosa}
\affiliation{Department of Applied Physics, The University of Tokyo, Hongo, 
Tokyo, 113-8656, Japan}
\affiliation{RIKEN Center for Emergent Matter Sciences (CEMS), Wako, Saitama, 
351-0198, Japan}

\begin{abstract}
Shift current is a photovoltaic current in bulk noncentrosymmetric insulator~\cite{Kraut1979,vBaltz1981,Sturman1992,Sipe2000,Tokura2018}. Studies on shift current have so far focused on the extended Bloch waves, such as in semiconductors~\cite{Sipe2000,Cook2017} and perovskites~\cite{Kraut1979,Young2012}. In contrast, it is unknown whether the localized wavefunctions support the dc photocurrent. Here, we show theoretically that the dc shift current appears in a noncentrosymmetric disordered one-dimensional insulator with random potential. When the light illuminates the entire sample, the photocurrent forms in presence of the electron-phonon coupling. We find this photocurrent remains robustly even when the energy scale of random potential is larger than the bandwidth. On the other hand, the photocurrent decays exponentially when the excitation is local, or the relaxation is only due to the contact with the electrodes. These results open a route to design high-efficiency solar cells and photodetectors.  
\end{abstract}

\date{\today}

\maketitle

\section*{Introduction}
Photovoltaic effects are the subject of vital importance both from the viewpoints of fundamental physics and applications. They are relevant to the solar cells and photodetectors. The basic process is that the photoexcited carriers, i.e., electrons and holes, are accelerated by the potential gradient or the electric field to produce the current. Therefore, it is necessary to separate the electron-hole pairs to free carriers.    
On the other hand, a photovoltaic effect of completely different mechanism has been proposed~\cite{Kraut1979,vBaltz1981,Sturman1992,Sipe2000,Tokura2018,Young2012,Cook2017}. It is caused by the geometrical nature of the electronic states in solids in noncentrosymmetric materials, which is characterized by the Berry phase.
 
Berry phase of electrons in solids governs many novel quantum transport phenomena~\cite{Xiao2010}, such as quantum Hall effect~\cite{Thouless1982}, anomalous Hall effect~\cite{Karplus1954,Nagaosa2010,Ishizuka2017b}, spin Hall effect~\cite{Murakami2003,Sinova2015}, magnetoresistance~\cite{Son2013}, and de Haas-van Alphen effect~\cite{Alexandradinata2018}. The basic idea is that the quantum states corresponding to a band form a manifold in Hilbert space, which are characterized by connection and curvature. Especially, the Berry connection has a physical meaning of intracell coordinate~\cite{Blount1950,KingSmith1993,Resta1993}. Therefore, it is relevant to low-energy processes, in which the electrons are confined in a band. However, it has been shown that Berry connection is also relevant to the phenomena involving interband transition. An example is the shift current, which is closely related to the ferroelectric polarization~\cite{vBaltz1981,Sipe2000}]; a shift of the intracell coordinate during the photo-excitation forms a photovoltaic current. In other words, the photo excitation of electrons changes the electric polarization, i.e., the polarization current, resulting in the dc current under the steady photoexcitation.

\begin{figure*}
  \includegraphics[width=0.9\linewidth]{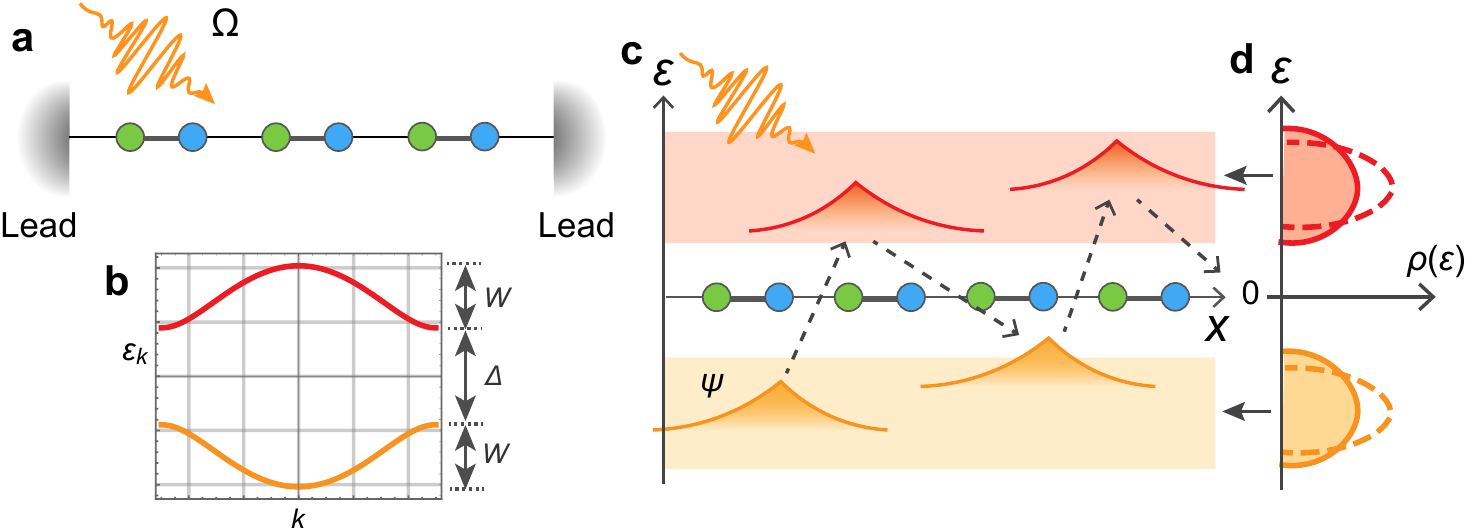}
  \caption{
  {\bf Schematic figures of the shift current in Anderson insulator.} {\bf a}. Schematic model of the system we consider; one-dimensional electron chain coupled to metal leads. {\bf b}. The band structure of this model in the clean limit, $V_{\rm rnd}=0$. {\bf c}. Schematic picture of photocurrent in noncentrosymmetric semiconductor in Anderson insulator. The photo-induced hopping between the localized states in the valence (yellow) and conduction (red) bands form macroscopic current. {\bf d}. Schematic figure of the density of states of the 1d Anderson insulator for $V_{\rm rnd}\lesssim \Delta$. The density of states consists of two groups of localized states (solid lines) separated by an energy gap. Each island corresponds to the conduction (red) and valence (yellow) bands in {\bf b} when $V_{\rm rnd}=0$. When $V_{\rm rnd}>0$, the impurity broadens the bands in the clean limit (dashed lines in {\bf d}). However, the energy gap remains if the disorder is weaker than the band gap, i.e., $V_{\rm rnd}\lesssim \Delta$. 
  }\label{fig:model}
\end{figure*}

In contrast to the extended Bloch wavefunctions, the wavefunctions are localized when the static disorder potential is strong enough, i.e., Anderson insulator~\cite{Kramer1993}. This qualitative difference prohibits the application of theories for clean materials to Anderson insulators. In the present paper, we study theoretically the impact of the localization on the shift current. Using Keldysh Green's function method, we evaluate the photocurrent in a one-dimensional (1d) chain attached to metal leads (figure~\ref{fig:model}a). The results show that the photocurrent forms in bulk Anderson insulator when the electrons couple to phonons. Intuitively, the current is a consequence of the optical transition of electrons between the localized valence- and conduction-band states, as shown in figure~\ref{fig:model}c. We also investigate the non-local nature of shift current~\cite{Ishizuka2017,Nakamura2017,Ogawa2017,Bajpai2019}; the photocurrent vanishes when the distance between the excited region and the leads is larger than the localization length. Our results show that Anderson insulators are a candidate for high-efficiency solar cells and optical sensors.

\section*{Results}

We consider a one-dimensional chain coupled to heat bath and Einstein phonons (figure~\ref{fig:model}a). The Hamiltonian for fermion chain reads
\begin{align}
H_0=&-t\sum_{i=1}^N[1+(-1)^i\delta]\hat c_{i+1}^\dagger \hat c_i+{\rm h.c.}
+\sum_i[V_{\rm s}(-1)^i+V_i]\hat c_i^\dagger \hat c_i,
\end{align}
where $\hat c_i$ ($\hat c_i^\dagger$) is the annihilation (creation) operator of the fermion at site $i$, $t$ is the hopping integral, $V_{\rm s}$ is the staggered potential and $V_i$ is the random potential at site $i$. We introduce the staggered hopping by $\delta$. The random potential $V_i$ has a uniform distribution of $V_i\in[-V_{\rm rnd},V_{\rm rnd}]$. In the following, we focus on the case $t=\sqrt{3/2}$, $\delta=1/\sqrt{3}$ and $V_{\rm s}=1/2$. In the clean limit $V_{\rm rnd}=0$, this model has one valence and one conduction bands, as shown in figure~\ref{fig:model}b and the dashed lines in figure~\ref{fig:model}d. The bandwidth and the energy gap between the valence and conduction bands are respectively $W=1$ and $\Delta=3$ in the calculation below.

In the presence of a weak disorder $V_{\rm rnd}\ll \Delta$, the density of states shows two islands that corresponds to the valence and conduction bands (solid lines in figure~\ref{fig:model}d). The eigenstates in these two bands are localized for arbitrary $V_{\rm rnd}>0$ because 1d electron systems are susceptible to the random potentials~\cite{Kramer1993}. We confirmed the localization using inverse participation ratio (Supplementary Information). The two-band model with a weak disorder is the situation we mainly consider in this work.

\begin{figure*}
  \includegraphics[width=\linewidth]{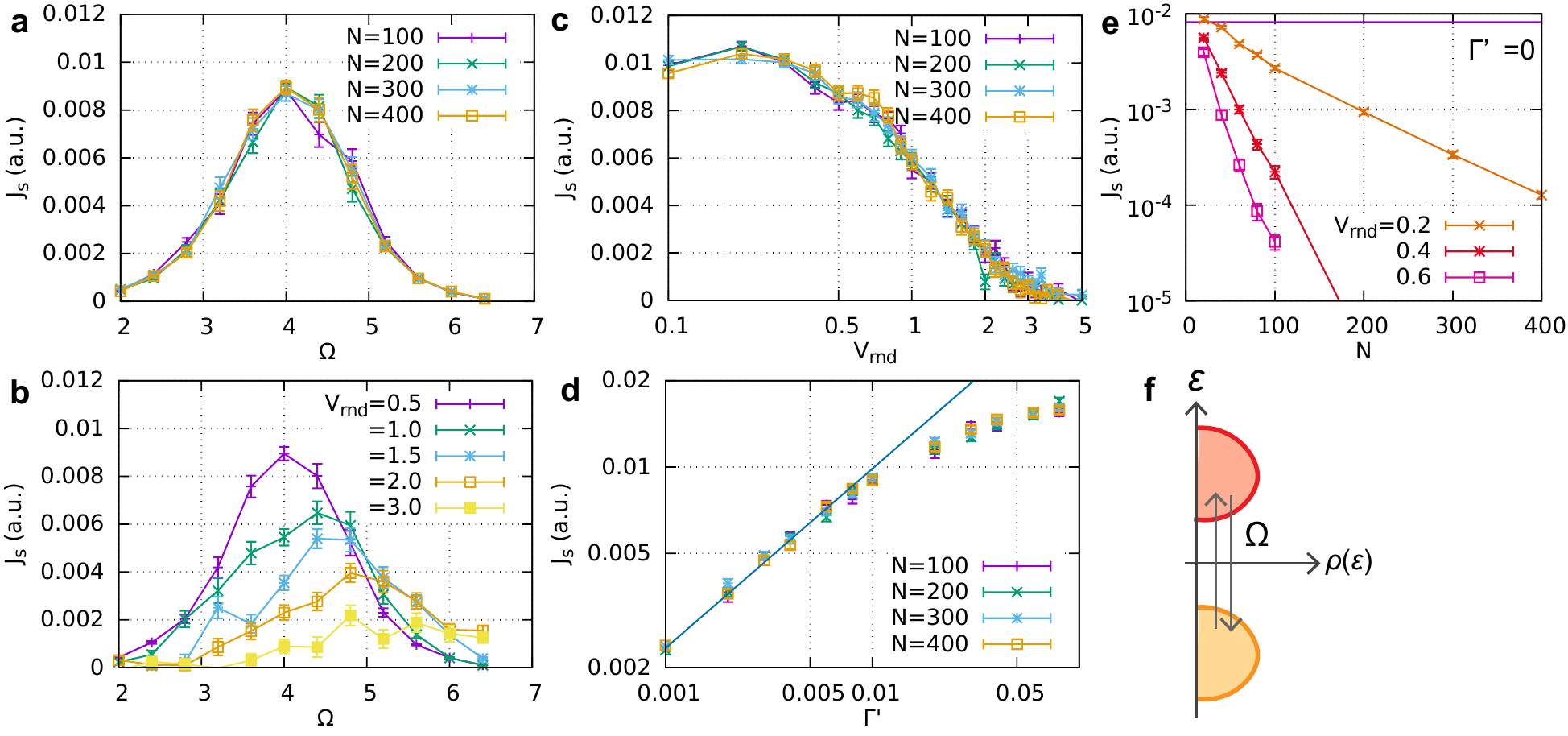}
  \caption{
  {\bf Photocurrent in the Anderson insulator with electron-phonon interaction.} {\bf a}. Light frequency $\Omega$ dependence of photocurrent. Different lines are for different length $N$ of the chain. {\bf b}. $\Omega$ dependence of photocurrent for different magnitude of random potential $V_{\rm rnd}$ ($N=400$). {\bf c}. $V_{\rm rnd}$ dependence of photocurrent for $\Omega=4$. {\bf d}. Electron-phonon interaction $\Gamma'$ dependence of photocurrent for $\Omega=4$. {\bf e}. Photocurrent $J_{\rm s}$ for chain without electron-phonon interaction ($\Gamma'=0$). {\bf f}. Schematic picture of the optical transition responsible for the photocurrent.
  }\label{fig:Js}
\end{figure*}

We calculate the photocurrent in the Anderson insulator by applying a Keldysh Green's function method~\cite{Haug2008}. The electric current we discuss is the current that flows from the chain to the two metal leads~\cite{Haug2008}. The coupling to the leads is expressed as an imaginary self-energy ${\rm i}\Gamma$ at the ends of the system~\cite{Meir1992}. In addition to the metal leads, we consider a similar dissipation term with coefficient ${\rm i}\Gamma'$ that couples to each site (see Method Section). This term mimics the dissipation by phonons within the first Born approximation.

We first show the $N$ dependence of the photocurrent for $\Gamma'=0$ in figure~\ref{fig:Js}e. The results are the average photocurrent for 200 different $V_i$ configurations. The transverse solid line is the result of photocurrent for $V_{\rm rnd}=0$ with $N=400$. By introducing a weak disorder, the current decays exponentially with $N$. This result is consistent with the IPR, which suggests $\xi\lesssim100$ sites for $V_{\rm rnd}\ge0.2$ (Supplementary Information). The result implies that the photocurrent vanishes in the bulk limit if no electron-phonon coupling exists.
 
The results are qualitatively different when the electron-phonon coupling exists ($\Gamma'\ne0$). Figure~\ref{fig:Js}a-\ref{fig:Js}d shows the numerical result of the photocurrent for $V_{\rm rnd}\ne0$. $\Omega$ dependence of the photocurrent for $V_{\rm rnd}=0.5$, $\Gamma=0.1$, $\Gamma'=0.01$ and $A=0.2$ is shown in figure~\ref{fig:Js}a. The results are the average of $J_{\rm s}$ for 200 different $V_i$ configurations. The four sets of data show the results for different $N$. The result shows the finite size effect is similar to or smaller than the statistical error; the result in the $N\to\infty$ limit should look like that for $N=400$. The result shows a finite photocurrent appears in a window of frequency $2\lesssim\Omega\lesssim6$. The window resembles that of the Bloch wavefunctions (clean limit)~\cite{Ishizuka2017}, which requires a direct transition between the conduction and valance bands. Indeed, the window corresponds to that of direct transition $\Omega\in[\Delta-2V_{\rm rnd},\Delta+2(W+V_{\rm rnd})]=[2,6]$ (figure~\ref{fig:Js}f). The result shows that the photocurrent remains finite in the $N\to\infty$ limit when $\Gamma'\ne0$, in contrast to the case without phonons.

The photocurrent robustly remains in the wide range of $V_{\rm rnd}$. Figure~\ref{fig:Js}b shows the $\Omega$ dependence of photocurrent for different $V_{\rm rnd}$. A finite photocurrent remains for $2V_{\rm rnd}>W=1$, i.e., when the random potential is larger than the bandwidth. Besides the suppression, peak $\Omega$ increases with increasing $V_{\rm rnd}$.  $V_{\rm rnd}$ dependence of the photocurrent for $\Omega=4$ is shown in figure~\ref{fig:Js}c. The current remains nearly constant for $2V_{\rm rnd}<W$. On the other hand, the current decreases for a larger $V_{\rm rnd}$ and strongly suppressed when $V_{\rm rnd}\gtrsim \Delta$. The suppression is a consequence of the two effects: the shift of the peak in the $\Omega$ dependence and the suppression of the current (see figure~\ref{fig:Js}b). The result shows that the photocurrent remains robust against the localization up to $V_{\rm rnd}\gtrsim\Delta$ when the electrons couple to phonons.
 
From the results, we find $J_s\sim(\Gamma')^{0.625\pm0.030}$ in the $\Gamma'/t\ll1$ limit. Figure~\ref{fig:Js}d shows the $\Gamma'$ dependence of $J_{\rm s}$ for different $N$. The result shows a negligible finite size effect for all ranges of $\Gamma'$. The solid line in the figure is the best fit of $N=400$ data with $f(x)=a (\Gamma')^\beta$; the best fit is $a=0.176\pm0.029$ and $\beta=0.635\pm0.030\sim2/3$. The result shows $J_{\rm s}$ is a sublinear function of $\Gamma'$ with a power close to 2/3. The result implies $J_s\propto T^{2/3}$ because $\Gamma'\propto T$ for the electron-phonon coupling in the high-temperature region.

The necessity of phonons is supposedly a manifestation of the role of dissipation. Theoretically, shift current is related to the causality, i.e., how we go around the poles in the denominator in the nonlinear response formula~\cite{vBaltz1981}. The imaginary part of the denominator is related to the dissipation, such as the imaginary part of Green's function. For example, the photocurrent vanishes in a clean Rice-Mele chain without dissipation~\cite{Morimoto2016}. In a clean metal, the leads introduce dissipation to all electron states because the electrons are extended. In contrast, in Anderson insulators, the leads only affect the states close to the edges because the localized wavefunctions deep in bulk are unaffected by the details of the surface. The electron-phonon coupling introduces the dissipation to these localized states. Therefore, the dissipation by phonons is essential for the generation of the photocurrent in Anderson insulators.

\begin{figure}
  \includegraphics[width=\linewidth]{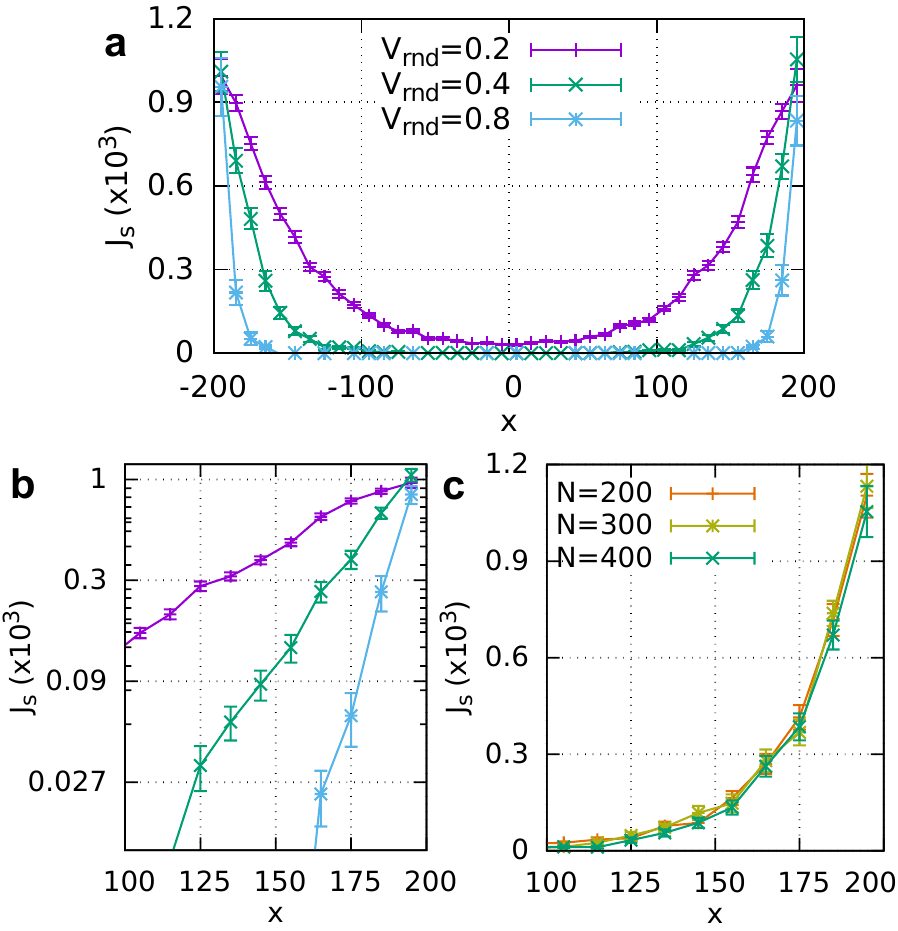}
  \caption{
  {\bf Shift current by local excitation.} {\bf a}. Light position $x$ dependence of the photocurrent. The results are for $N=400$, $\Omega=4.0$, $\Gamma'=0.01$, and $\Gamma=0.1$. {\bf b}. Logarithmic plot of the right end of the data in {\bf a}. {\bf c}. The similar results to panel {\bf a.} near the right end of the sample with $V_{\rm rnd} = 0.4$ for different sample sizes, $N=100$, $200$, $300$, and $400$.
  }\label{fig:local}
\end{figure}

We next look at the local excitation of the photocurrent. In figure~\ref{fig:local}, we shine the light on to $l=10$ sites of the chains at position $x$, i.e., the bonds connecting sites $i=x-(l-1)/2,x-(l-1)/2+1,\cdot,x+(l-1)/2$ are subject to the light. The position dependence of the photocurrent for $N=400$ is shown in figure~\ref{fig:local}a. $x=\pm200$ corresponds to the case in which the light position is at the ends of the chain and $x=0$ is when the light is at the center. Unlike the case in the clean limit~\cite{Ishizuka2017,Nakamura2017,Ogawa2017}, the current decays when the light position goes away from the two ends. Figure~\ref{fig:local}b is the logarithmic plot of the $x$ dependence of $J_{\rm s}$ near the right end of figure~\ref{fig:local}a. The current decays exponentially with respect to $x$. Figure~\ref{fig:local}c shows the decay of the current near the right end for different sample size $N$. The result shows negligible size dependence. The result shows the current decays exponentially with respect to the position of the light, qualitatively different from the clean limit.

\section*{Disussion}

Our numerical study on the disordered fermion chain shows the localized electrons in Anderson insulator form photovoltaic current. An important feature here is the electron-phonon coupling, which we introduce within the first Born approximation. Our result shows the photocurrent remains finite in the $N\to\infty$ limit if the weak electron-phonon coupling exists. This feature is in contrast to the case without phonons. In this case, the photocurrent vanishes once the length of the chain exceeds the localization length of the wavefunction. The result shows that Anderson insulators function as a solar cell.

Our results are relevant to noncentrosymmetric disordered semiconductors, in which the carriers are in the Anderson-localized states. Our results predict a robust photovoltaic current in the disordered semiconductors, which is proportional to $J_{\rm s}\propto(\Gamma')^{\frac23}$. This result implies $J_{\rm s}\propto T^{\frac23}$ if $\Gamma'$ arises from the electron-phonon coupling. In contrast, the Ohmic conductivity follows $\sigma\propto e^{-(T_0/T)^\beta}$, where $\beta=1/2$ in the case of 1d Anderson insulator~\cite{Mott1969,Apsley1974}. Therefore, the Ohmic conductivity decreases faster than the photocurrent conductivity as we decrease the temperature. The suppression of the Ohmic dc current contributes to the efficiency of solar cells by reducing the current back-flow. For instance, a recent work discusses that the shift current device acts as a current source; the output current reads $J_{\rm out}=J_{\rm s}R^\ast/(2R_0+R^\ast)$ where $R^\ast$ and $R_0$ are respectively the resistivity of the photovoltaic device and the leads~\cite{Nakamura2020}. The result implies materials with high $R^\ast$ are beneficial for applications. Therefore, disordered semiconductors are an interesting candidate for high-efficiency solar cell.

\section*{Methods}

We consider an electron chain
\begin{align}
H=H_0+H_{\rm p}+H_{\rm ep},
\end{align}
where
\begin{align}
H_{\rm p}=\sum_q\omega_q a_q^\dagger a_q,
\end{align}
is the phonon Hamiltonian and
\begin{align}
H_{\rm ep}=\sum_{n,k,q} M_q c_{n,k+q}^\dagger c_{n,k}(a_q+a_{-q}^\dagger),
\end{align}
is the electron-phonon interaction. Here, $c_{n,k}$ ($c_{n,k}^\dagger$) and $a_k$ ($a_k^\dagger$) are respectively the electron and phonon annihilation (creation) operators, $\omega_q$ is the phonon eigenenergy with momentum $q$, and $M_q$ is the electron-phonon coupling. $H_0$ is defined in the main text.

We used Keldysh Green's function method for the calculation of photocurrent in the main text. The Green's function for the nonequilibrium state is obtained by numerically diagonalizing the Dyson equation~\cite{Haug2008},
\begin{align}
  (\omega+n\Omega)G_{n,n'}(\omega)&- \sum_{n''}{\cal H}_{n,n''}^0\, G_{n'',n'}(\omega)\nonumber\\
  &- \sum_{n''} \Sigma'_{n,n''}(\omega)\,G_{n'',n'}(\omega) = 1.\label{eq:dyson}
\end{align}
Here, $G_{n,n'}(\omega)$ is the $2N\times2N$ square matrix of Keldysh Green's function, ${\cal H}_{n,n'}^0=\frac1T\int_0^T dt H_0(t)\,e^{\mathrm i (n-n') \Omega t}$ is the Fourier transform of the Hamiltonian in the maintext, and $\Sigma'_{n,n''}(\omega)$ is the self energy.

In our calculation, the coupling to the leads and to the phonons are taken into account as the self energy. In the main text, we consider the self energy of form
\begin{align}
  &\left[ \Sigma'_{\kappa,\kappa}(\omega) \right]_{i,j}=\nonumber\\
  &\left[{\rm i}\Gamma(\omega+\kappa\Omega)\left\{\delta_{i,0}\delta_{j,0}+\delta_{i,N-1}\delta_{j,N-1}\right\}+{\rm i}\Gamma'(\omega+\kappa\Omega)\right]\nonumber\\
  &\left( 
  \begin{array}{cc}
  -\frac12 & 2f(\omega+\kappa\Omega)-1 \\
  0 & \frac12
  \end{array}
  \right).
  \label{eq:Sigma}
\end{align}
Here, $\Gamma(\omega)$ is the self-energy reflecting the coupling to the leads~\cite{Meir1992},
\begin{align}
  \Gamma(\omega) = 2\pi \sum_l |V_{l,0}|^2 \delta(\omega-\varepsilon_{l})=2\pi \sum_l |V_{l,N-1}|^2 \delta(\omega-\varepsilon_{l}),
\end{align}
where $\varepsilon_{k\sigma}$ is the eigenenergy of the lead with index $l$ and $V_{l,i}$ is the hopping integral between the electrons on site $i$ and the lead state $l$. We assume the coupling is the same for left and right leads. In addition to the leads, we consider electron-phonon interaction represented by $\Gamma'(\omega)$. Within first Born approximation, $\Gamma'(\omega)$ reads~\cite{Haug2008}
\begin{align}
\Sigma^{\rm(ph)}(k,\omega)=\sum_{q}\int\frac{d\omega'}{2\pi}M_q^2G^0_{k-q}(\omega-\omega')D^0_q(\omega'),
\end{align}
where $G^0_{k}(\omega)$ is the unperturbed Green function for $H_0$, and $D^0_q(\omega)$ is the free phonon Green's function. We here approximate $G^0_k(\omega)$ and $D^0_q(\omega)$ by that for periodic system. With these approximations, the imaginary part of the self energy reads
\begin{align}
\Sigma^{\rm(ph)}_{\kappa,\kappa}(k,\omega)=
  {\rm i}\Gamma'(\omega+\kappa\Omega)\left( 
  \begin{array}{cc}
  -\frac12 & 2f(\omega+\kappa\Omega)-1 \\
  0 & \frac12
  \end{array}
  \right),
\end{align}
where
\begin{align}
\Gamma'(\omega)\sim2\pi\sum_q|M_q|^2(2N_q+1)\delta(\omega-\varepsilon_{k-q}),
\end{align}
$N_q$ and $M_q$ are the number of phonons and the electron-phonon coupling between the electrons and the phonon with momentum $q$. In the above $\Gamma'$, we assumed the eigenenergy of the phonons are much smaller than that of electrons $|\varepsilon_{k}|$.

For simplicity, we assume $\Gamma$ and $\Gamma'$ to be a constant of $\omega$.

\begin{acknowledgements}
We thank enlightening discussions with M. Kawasaki, T. Morimoto, Y. Nakamura, N. Ogawa, M. Sotome, and Y. Tokura. This work was supported by JST CREST Grant Numbers JPMJCR1874 and JPMJCR16F1, Japan, and JSPS KAKENHI Grant Numbers JP18H03676, JP18H04222, JP19K14649 and JP26103006. HI was partly supported by UTokyo Global Activity Support Program for Young Researchers.
\end{acknowledgements}

\end{document}